\documentclass[12pt]{article}

\usepackage{graphicx}
\usepackage{float}

\def\gtwid{\mathrel{\raise.3ex\hbox{$>$\kern-.75em\lower1ex\hbox{$\sim$}}}}
\def\ltwid{\mathrel{\raise.3ex\hbox{$<$\kern-.75em\lower1ex\hbox{$\sim$}}}}
\def\square{\kern1pt\vbox{\hrule height 1.2pt\hbox{\vrule width 1.2pt\hskip 3pt
   \vbox{\vskip 6pt}\hskip 3pt\vrule width 0.6pt}\hrule height 0.6pt}\kern1pt}

\begin{document}

\begin{titlepage}

\begin{flushright}
UFIFT-QG-18-04
\end{flushright}

\vskip 2cm

\begin{center}
{\bf Cosmological Coleman-Weinberg Potentials and Inflation}
\end{center}

\vskip 1cm

\begin{center}
J. H. Liao$^{1*}$, S. P. Miao$^{1\star}$ and R. P. Woodard$^{2\dagger}$
\end{center}

\vskip .5cm

\begin{center}
\it{$^{1}$ Department of Physics, National Cheng Kung University \\
No. 1, University Road, Tainan City 70101, TAIWAN}
\end{center}

\begin{center}
\it{$^{2}$ Department of Physics, University of Florida,\\
Gainesville, FL 32611, UNITED STATES}
\end{center}

\vspace{1cm}

\begin{center}
ABSTRACT
\end{center}
We consider an additional fine-tuning problem which afflicts scalar-driven
models of inflation. The problem is that successful reheating requires the 
inflaton be coupled to ordinary matter, and quantum fluctuations of this
matter induces Coleman-Weinberg potentials which are not Planck-suppressed. 
Unlike the flat space case, these potentials depend upon a still-unknown, 
nonlocal functional of the metric which reduces to the Hubble parameter 
for de Sitter. Such a potential cannot be completely subtracted off by any 
local action. In a simple model we numerically consider one possible 
subtraction scheme in which the correction is locally subtracted at the 
beginning of inflation. For fermions the effect is to make the universe
approach de Sitter with a smaller Hubble parameter. For gauge bosons the 
effect is to make inflation end almost instantly unless the gauge charge 
is unacceptably small.
 
\begin{flushleft}
PACS numbers: 04.50.Kd, 95.35.+d, 98.62.-g
\end{flushleft}

\begin{flushleft}
$^{*}$ e-mail: a0983028669@gmail.com \\
$^{\star}$ e-mail: spmiao5@mail.ncku.edu.tw \\
$^{\dagger}$ e-mail: woodard@phys.ufl.edu
\end{flushleft}

\end{titlepage}

\section{Introduction}

The most recent results for the scalar spectral index $n_s$, and the limits
on the tensor-to-scalar ratio $r$ \cite{Ade:2015lrj}, are still consistent 
with certain models of single scalar-driven inflation,
\begin{equation}
\mathcal{L} = \frac{R \sqrt{-g}}{16 \pi G} - \frac12 \partial_{\mu} \varphi
\partial_{\nu} \varphi g^{\mu\nu} \sqrt{-g} - V(\varphi) \sqrt{-g} \; . 
\label{GRMCS}
\end{equation}
However, the allowed models suffer from severe fine-tuning problems associated 
with the need to keep the potential very flat, with getting inflation to start
and with avoiding the loss of predictivity through the formation of a multiverse 
\cite{Ijjas:2013vea}. This has led to much controversy within the inflation 
community \cite{Guth:2013sya,Linde:2014nna,Ijjas:2014nta}. 

The purpose of this paper is to study a different sort of fine-tuning problem 
which is associated with the necessity of coupling the inflaton to normal 
matter to make reheating efficient. It has long been known that the quantum 
fluctuations of such matter particles will induce Coleman-Weinberg corrections 
to the inflaton effective potential \cite{Coleman:1973jx}. These corrections 
are dangerous for inflation because they are not Planck-suppressed 
\cite{Green:2007gs}.

Until recently the assumption was that cosmological Coleman-Weinberg potentials
are simply local functions of the inflaton which could be subtracted at will.
However, existing results (from scalars \cite{Janssen:2008px}, from fermions 
\cite{Candelas:1975du,Miao:2006pn}, and from gauge bosons \cite{Allen:1983dg,
Prokopec:2007ak}) on de Sitter background show that the corrections actually 
take the form of the fourth power of the Hubble constant times a complicated 
function of the dimensionless ratio of the inflaton to the Hubble constant. 
Simple arguments show that these factors of the de Sitter Hubble parameter 
cannot be constant for evolving cosmologies, and are not even local functionals 
of the metric \cite{Miao:2015oba}. Of course this means that they cannot be 
completely subtracted.

In this paper we study one possible partial subtraction scheme. Because 
cosmological Coleman-Weinberg potentials are only known for de Sitter we shall
make the {\it Instantaneous Hubble Approximation} in which the de Sitter Hubble 
constant is replaced by the evolving Hubble parameter. Our scheme is to subtract 
the same term with the Hubble parameter evaluated at the initial time, so that 
the cancellation is perfect at the initial time. Section 2 of this paper explains 
why very weak matter couplings are disfavored. The appropriate modified Friedmann 
equations are derived in section 3. In section 4 we study the effects of potentials 
induced by fermions and by gauge bosons. Section 5 presents our conclusions.

\section{Connecting Reheating and Fine Tuning}

The universe must reheat before the onset of Big Bang Nucleosynthesis but 
this seeming lower bound can only be achieved through a high degree of fine
tuning. Simple models of inflation all require much higher reheat temperatures. 
Given any model one can use the observed values of the scalar amplitude $A_s$ 
and the scalar spectral index $n_s$ to compute both the number of e-foldings 
from when observable perturbations experienced first horizon crossing to now, 
and also the number of e-foldings from 1st crossing to the end of inflation. 
The difference between these two is the number of e-foldings from the end of 
inflation to now, during which reheating must occur. For example, in the 
$V = \frac12 m^2 \varphi^2$ model we will study, the difference is 
\cite{Mielczarek:2010ag},
\begin{equation}
\Delta N = \frac12 \ln\Biggl[ \frac{\pi (1 - n_s) A_s}{G k_0^2}\Biggr] -
\frac2{1 - n_s} \; , \label{DeltaN1}
\end{equation}
where $k_0$ is the pivot wave number. With 2015 Planck numbers \cite{Ade:2015lrj}
this works out to be about $\Delta N \simeq 61.3$ e-foldings.

The number of e-foldings since the end of inflation can be computed
independently and it has long been known to depend on the reheat temperature 
$T_R$ like $-\frac13 \ln(T_R)$. For example, the $V = \frac12 m^2 \varphi^2$
model gives \cite{Mielczarek:2010ag},
\begin{equation}
\Delta N = \frac13 \ln\Biggl[ \frac{15 (1 - n_s)^2 A_s}{128 \pi^2 G^2 T_{R}
T_0^3}\Biggr] \simeq 63.9 - \frac23 \ln(G T^2_R) \; , \label{DeltaN2}
\end{equation}
where $T_0$ is the current temperature of the cosmic microwave 
radiation.\footnote{Note the interesting fact that the number of relativistic 
species at the end of inflation drops out of this result.} The reason that high 
reheat temperatures are favored is that continuations of the simple models which 
describe the observed power spectrum correspond to small values of $\Delta N$,
which requires large $T_{R}$. For example, equating (\ref{DeltaN1}) and 
(\ref{DeltaN2}) implies a trans-Planckian reheat temperature! Of course the
uncertainties on $T_R$ are great owing to the exponential dependence on the
factor of $\frac2{1 - n_s}$ in (\ref{DeltaN1}), but the preference for large 
reheat temperatures is clear.

Considering more general models in the context of WMAP data, Martin and Ringeval 
derived a lower bound of more than $10^4~{\rm GeV}$ \cite{Martin:2010kz}. These 
results can only be evaded by decreasing the number of e-foldings between first
crossing and the end of inflation, which requires tuning the lower portion of the 
inflaton potential to be steeper than the portion during which observable 
perturbations experience first crossing. That raises obvious questions about why 
the potential changed form, and why the initial condition was such that observable 
perturbations happened to be generated when the scalar was on the flat portion. 

The preceding considerations were purely geometrical and had nothing to do with
specific mechanisms of reheating. We shall consider two matter couplings between
real and complex inflatons $\varphi$,
\begin{equation}
\Delta \mathcal{L}_1 = - \lambda \varphi \overline{\psi} \psi \sqrt{-g} \quad ,
\quad \Delta \mathcal{L}_2 = - \Bigl(\partial_{\mu} - i q A_{\mu}\Bigr) \varphi
\Bigl(\partial_{\nu} + i q A_{\nu}\Bigr) \varphi^* g^{\mu\nu} \sqrt{-g} \; .
\label{couplings}
\end{equation}
In the $V = \frac12 m^2 \varphi^2$ model inflation ends with an approximately 
matter dominated phase during which the scalar oscillates as energy gradually 
drains from it into ordinary matter through one or the other of the couplings 
(\ref{couplings}). With the $\Delta \mathcal{L}_1$ coupling the inflaton decays 
into two fermions at a rate of $\Gamma = \frac{\lambda^2 m}{8\pi}$. Reheating 
ends when the Hubble parameter falls below this rate and the reheat temperature 
can be estimated as \cite{Kofman:1997yn},
\begin{equation}
T_R \simeq \frac15 \Bigl( \frac{\Gamma^2}{G}\Bigr)^{\frac14} \simeq \lambda
\times 10^{15}~{\rm GeV} \; . \label{Treheat}
\end{equation}
With the $\Delta \mathcal{L}_2$ coupling the mechanism of reheating is through
parametric resonance \cite{Kofman:1997yn}. Estimating the reheat temperature 
requires numerical analysis but it is known that the process cannot be efficient 
for very small couplings $q^2 \ll 1$ \cite{Greene:1997fu}. 

\section{The Modified Friedmann Equations}

The purpose of this section is to work out how the Friedmann equations
change when the scalar potential is allowed to depend on the Hubble parameter,
$V(\varphi) \longrightarrow V(\varphi,H)$. Our technique exploits the famous
theorem \cite{Palais:1979rca,Torre:2010xa} that specializing to a class of
geometries before varying the action gives correct equations, even though it
can miss constraints. The restriction to homogeneity and isotropy give the
$ij$ Einstein equation and the scalar evolution equation, from which we infer
the $00$ equation. We then reduce these three equations to a dimensionless form.

We know the scalar potential model Lagrangian (\ref{GRMCS}) for arbitrary 
metric and scalar field configurations $g_{\mu\nu}(t,\vec{x})$ and 
$\varphi(t,\vec{x})$. This makes it simple to vary the action {\it first} and
then specialize to homogeneity and isotropy,
\begin{equation}
ds^2 = -dt^2 + a^2(t) d\vec{x} \!\cdot\! d\vec{x} \qquad , \qquad \varphi =
\varphi_0(t) \; . \label{homoiso}
\end{equation}
The two nontrivial Einstein equations are the $00$ and $ij$ components,
\begin{eqnarray}
3 H^2 & = & 8\pi G \Bigl[\frac12 \dot{\varphi}_0^2 + V(\varphi_0)\Bigr] \; ,
\label{G00} \\
-2 \dot{H} - 3 H^2 & = & 8\pi G \Bigl[ \frac12 \dot{\varphi}_0^2 - V(\varphi_0)
\Bigr] \; . \label{Gij}
\end{eqnarray}
The scalar equation is,
\begin{equation}
\ddot{\varphi}_0 + 3 H \dot{\varphi}_0 + \frac{\partial V}{\partial \varphi_0}
= 0 \; . \label{phieqn}
\end{equation}
Note the close relation which exists between the three equations,
\begin{equation}
\frac{d}{dt} \Bigl[ {\rm Eqn} (\ref{G00}) \Bigr] + 3 H \Bigl[ {\rm Eqn} (\ref{G00}) 
+ {\rm Eqn} (\ref{Gij})\Bigr] = 8 \pi G \, \dot{\varphi}_0 \Bigl[ {\rm Eqn} 
(\ref{phieqn}) \Bigr] \; . \label{usualrelation}
\end{equation}

Even with the replacement $H_{\rm dS} \longrightarrow H(t)$ in our de Sitter
results for Coleman-Weinberg potentials we still do not know how the Lagrangian 
depends upon a general field configuration. What we know is its specialization 
to homogeneity and isotropy (\ref{homoiso}) {\it before} variation,
\begin{equation}
L = \frac12 a^3 \dot{\varphi}_0^2 - a^3 V(\varphi_0,H) - \frac{6 a^3 H^2}{16 \pi G}
\; . \label{Lhomoiso}
\end{equation}
This might be thought to be a debilitating problem but it is not. We simply 
appeal to the theorem of Palais \cite{Palais:1979rca,Torre:2010xa} that all the
equations arising from such a specialized Lagrangian are at least correct, even 
though there may be additional equations. The Euler-Lagrange equation for 
$\varphi_0(t)$ is identical to (\ref{phieqn}). The Euler-Lagrange equation
for $a(t)$ follows from the derivatives of (\ref{Lhomoiso}) with respect to $a$ 
and $\dot{a}$,
\begin{eqnarray}
\frac{\partial L}{\partial a} & = & \frac{6 a^2}{16 \pi G} \Biggl\{ 8\pi G \Bigl[
\frac12 \dot{\varphi}_0^2 - V(\varphi_0,H) + \frac13 H 
\frac{\partial V(\varphi_0,H)}{\partial H} \Bigr] - H^2 \Biggr\} \; , 
\label{dLda} \\
\frac{\partial L}{\partial \dot{a}} & = & -\frac{6 a^2}{16 \pi G} \Biggl\{ 8\pi G 
\Bigl[ \frac13 \frac{\partial V(\varphi_0,H)}{\partial H} \Bigr] + 2 H\Biggr\} \; .
\label{dLdadot}
\end{eqnarray}
Hence we arrive at the appropriate generalization of equation (\ref{Gij}),
\begin{equation}
-2\dot{H} - 3 H^2 = 8\pi G \Bigl[ \frac12 \dot{\varphi}_0^2 - V +
H \frac{\partial V}{\partial H} + \frac13 \dot{\varphi}_0 
\frac{\partial^2 V}{\partial \varphi_0 \partial H} + \frac13 \dot{H}
\frac{\partial^2 V}{\partial H^2} \Bigr] \; . \label{newGij}
\end{equation}

The homogeneous and isotropic Lagrangian (\ref{Lhomoiso}) does not give us the
generalization of equation (\ref{G00}). However, we can guess it, guided by three
principles:
\begin{itemize}
\item{The generalization must reduce to (\ref{G00}) when the potential has no
dependence on $H$;}
\item{The generalization must not involve either $\ddot{\varphi}_0$ or 
$\ddot{a}$; and}
\item{Substituting the generalization for (\ref{G00}), and equation (\ref{newGij})
for (\ref{Gij}), in relation (\ref{usualrelation}) should give the scalar 
evolution equation.}
\end{itemize}
The desired generalization of (\ref{G00}) is easily seen to be,
\begin{equation}
3 H^2 = 8\pi G \Bigl[ \frac12 \dot{\varphi}_0^2 + V - H \frac{\partial V}{\partial H}
\Bigr] \; . \label{newG00}
\end{equation}

Relations (\ref{phieqn}), (\ref{newGij}) and (\ref{newG00}) define how the scalar
and the geometry of inflation evolve, but they are inconvenient because the scale
of temporal variation changes dramatically over the course of inflation, and because
the dependent variables are dimensionful. A more physically meaningful evolution
parameter is the number of e-foldings since the beginning of inflation, 
\begin{equation}
n \equiv \ln\Bigl[ \frac{a(t)}{a(t_i)}\Bigr] \qquad \Longrightarrow \qquad \frac{d}{dt} 
= H \frac{d}{d n} \quad , \quad \frac{d^2}{d t^2} = H^2 \Bigl[ \frac{d^2}{d n^2}
- \epsilon \frac{d}{d n}\Bigr] \; . \label{ndef}
\end{equation}
The natural dimensionless fields and potential are,
\begin{equation}
\phi(n) \equiv \sqrt{8 \pi G} \, \varphi_0(t) \; , \;
\chi(n) \equiv \sqrt{8 \pi G} \, H(t) \; , \;
U(\phi,\chi) \equiv (8 \pi G)^2 V(\varphi_0,H) \; . \label{dimless}
\end{equation}
With these changes, the modified Friedmann equations (\ref{newG00}) and (\ref{newGij})
take the form,
\begin{eqnarray}
3 \chi^2 & = & \frac12 \chi^2 {\phi'}^2 + U - \chi \frac{\partial U}{\partial \chi} 
\; , \label{F1} \\
-2 \chi \chi' - 3 \chi^2 & = & \frac12 \chi^2 {\phi'}^2 - U + \chi 
\frac{\partial U}{\partial \chi} + \frac13 \chi \phi' \frac{\partial^2 U}{\partial \phi
\partial \chi} + \frac13 \chi \chi' \frac{\partial^2 U}{\partial \chi^2} \; .
\label{F2} \qquad
\end{eqnarray}
And the scalar evolution equation becomes,
\begin{equation}
\phi'' + (3 \!-\! \epsilon) \phi' + \frac1{\chi^2} \frac{\partial U}{\partial \phi} 
= 0 \; , \label{F3}
\end{equation}
where the first slow roll parameter is,
\begin{equation}
\epsilon(n) \equiv -\frac{\chi'}{\chi} = \frac{ \frac12 {\phi'}^2 + \frac{\phi'}{6 \chi} 
\frac{\partial^2 U}{\partial \phi \partial \chi}}{1 + \frac16 \frac{\partial^2 U}{
\partial \chi^2}} \; . \label{epsilon}
\end{equation}
Finally, note that the leading slow roll approximations for the scalar and tensor power 
spectra take the form,
\begin{equation}
\Delta^2_{\mathcal{R}}(n) \approx \frac1{8 \pi^2} \times \frac{\chi^2(n)}{\epsilon(n)}
\qquad , \qquad \Delta^2_{h}(n) \approx \frac1{8 \pi^2} \times 16 \chi^2(n) \; .
\label{slowdelta}
\end{equation}

\section{The Fate of the $m^2 \varphi^2$ Model}

It is useful to study what Coleman-Weinberg corrections do to the familiar $V = \frac12 
m^2 \varphi^2$ model, even though that model is no longer consistent with the data. In
the slow roll approximation the evolution of the dimensionless scalar and the first
slow roll parameter are independent of the mass term,
\begin{equation}
{\rm Slow\ Roll} \quad \Longrightarrow \quad \phi(n) \simeq \sqrt{\phi^2(0) - 4 n} 
\quad , \quad \epsilon(n) \simeq \frac{2}{\phi^2(0) \!-\! 4n} \; . \label{slowroll}
\end{equation}
To make inflation last about 100 e-foldings (without the Coleman-Weinberg correction)
we choose the initial conditions,
\begin{equation}
\phi(0) = 20 \qquad , \qquad \phi'(0) = -\frac1{10} \; . \label{initial}
\end{equation}
We will continue using these conditions after the Coleman-Weinberg potential is added, 
with the initial value of $\chi$ chosen to obey equation (\ref{F1}). We parameterize 
the mass in terms of a constant $k$ which is chosen to make the amplitude of the 
scalar power spectrum agree with observation \cite{Ade:2015lrj} (again, without the 
Coleman-Weinberg correction),\footnote{The tensor-to-scalar ratio of $r \simeq 0.16$ 
does {\it not} agree with observation \cite{Ade:2015lrj}, which is why this model is 
disfavored. However, it is very simple, and well known, and the robustness of our 
results does not justify employing a more viable model.}
\begin{equation}
m^2 \equiv \frac{k^2}{8\pi G} \qquad , \qquad \frac{(202 k)^2}{96 \pi^2} \simeq 2 
\times 10^{-9} \; . \label{kdef}
\end{equation}
This defines the classical model which is being corrected. We first consider
an inflaton which is Yukawa-coupled to a fermion, then we consider a charged inflaton 
which is coupled to a gauge boson. In each case the Coleman-Weinberg potential has
disastrous consequences.

\subsection{Inflaton Yukawa-Coupled to Fermions}

If the Yukawa coupling constant is $\lambda$, and we subtract the quantum correction
at $n = 0$, the dimensionless potential is, 
\begin{equation}
U(\phi,\chi) = \frac12 k^2 \phi^2 - \frac{\chi^4}{8 \pi^2} f\Bigl( 
\frac{\lambda \phi}{\chi}\Bigr) + \frac{\chi^4(0)}{8\pi^2} f\Bigl( 
\frac{\lambda \phi}{\chi(0)}\Bigr) \; . \label{fermiU}
\end{equation}
Here the scalar dependent part of the Coleman-Weinberg potential is
\cite{Candelas:1975du,Miao:2006pn},
\begin{equation}
f(z) = 2 \gamma z^2 - \Bigl[ \zeta(3) \!-\! \gamma\Bigr] z^4 + 2 \int_{0}^{z} 
\!\!\! dx \, (x \!+\! x^3) \Bigl[ \psi(1 \!+\! i x) + \psi(1 \!-\! i x)\Bigr] 
\; , \label{fermif}
\end{equation}
where $\psi(x) \equiv \frac{d}{dx} \ln[\Gamma(x)]$ is the digamma function.
The small value of $k^2 \sim 4 \times 10^{-11}$ needed to reproduce the
scalar amplitude (\ref{kdef}) means that the quantum corrections tend to
overwhelm the classical term in (\ref{fermiU}), unless the Yukawa coupling
is chosen to be very small. With order one values of $\lambda$ there is no
evolution at all. This is because the middle term of (\ref{fermiU}) 
decreases relative the the final term as a function of $\chi$. Hence a
putative decrease in $\chi$ would actually {\it increase} $U(\phi,\chi)$,
which is inconsistent with equation (\ref{F1}), unless the classical term
dominates the two quantum corrections. 

\begin{figure}[H]
\includegraphics[width=4.0cm,height=3.0cm]{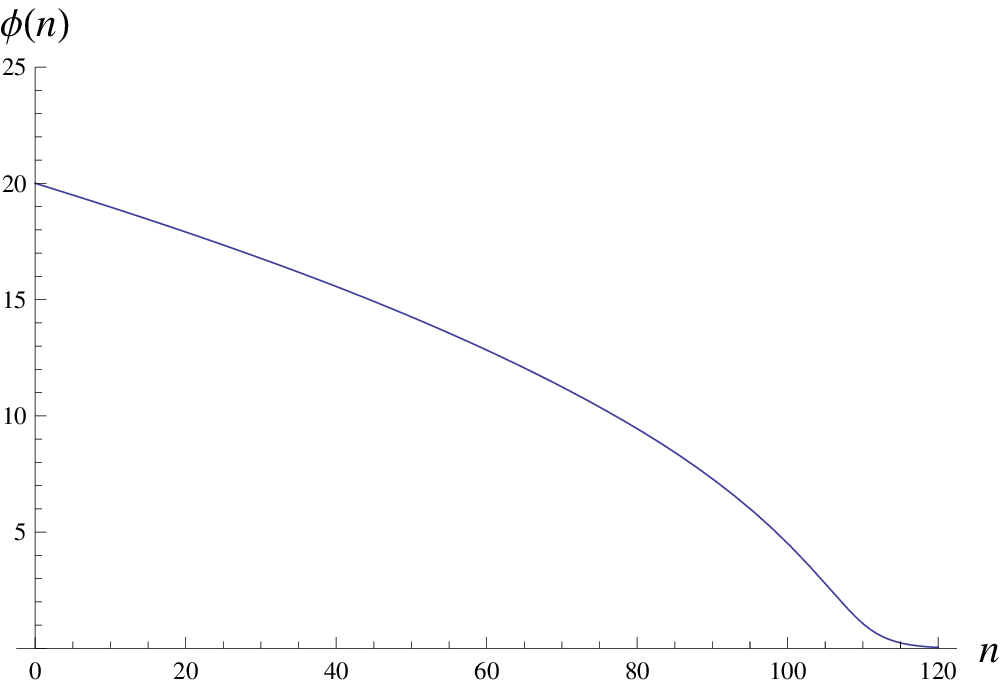}
\hspace{.5cm}
\includegraphics[width=4.0cm,height=3.0cm]{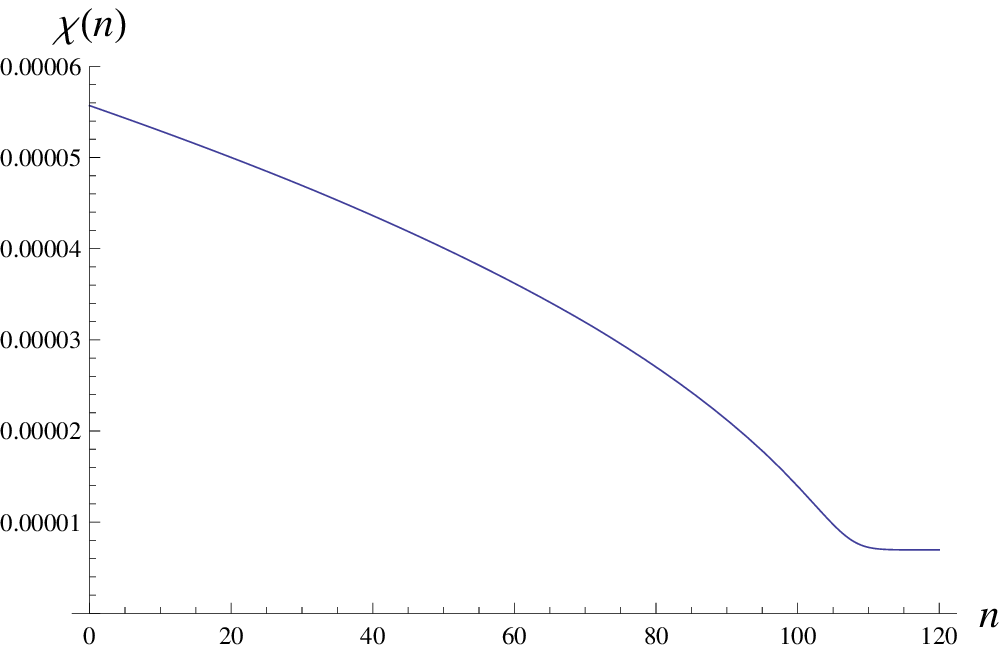}
\hspace{.5cm}
\includegraphics[width=4.0cm,height=3.0cm]{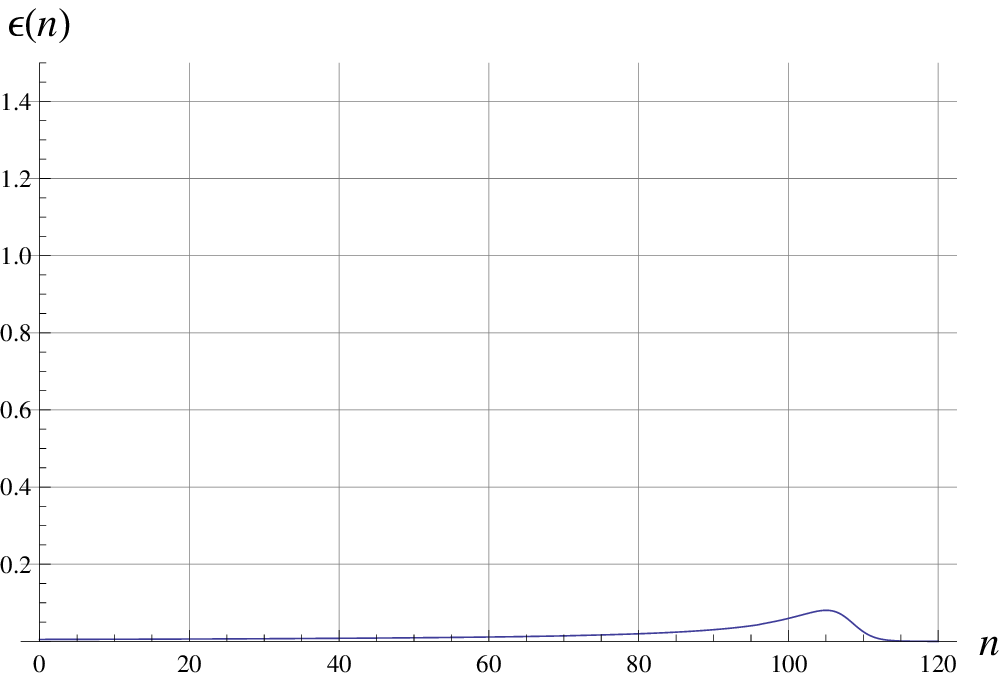}
\caption{Plots of the dimensionless scalar $\phi(n)$ (on the left), the
dimensionless Hubble parameter $\chi(n)$ (middle) and the first slow roll
parameter $\epsilon(n)$ (on the right) for the quantum-corrected model 
(\ref{fermiU}) with Yukawa coupling $\lambda = 5 \times 10^{-4}$. Even 
with this minuscule value of $\lambda$ the geometry approaches de Sitter
at a reduced scale.}
\label{fermifigs2}
\end{figure}

We did not start to see evolution until values of about $\lambda \sim 10^{-3}$. 
Figure~\ref{fermifigs2} shows the result for $\lambda = 5 \times 10^{-4}$. 
Although the model evolves noticeably for the first 100 e-foldings, there are 
considerable deviations from the classical result. These deviations become 
extreme at late times, for which the figure shows that the quantum-corrected 
model approaches de Sitter expansion at a reduced Hubble parameter.

\begin{figure}[H]
\includegraphics[width=4.0cm,height=3.0cm]{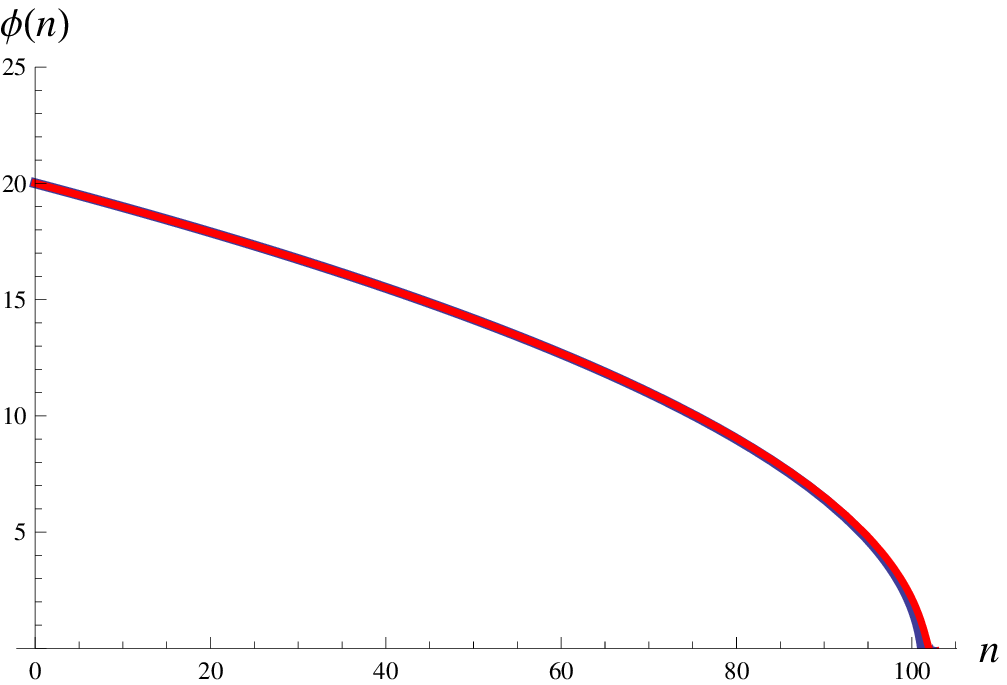}
\hspace{.5cm}
\includegraphics[width=4.0cm,height=3.0cm]{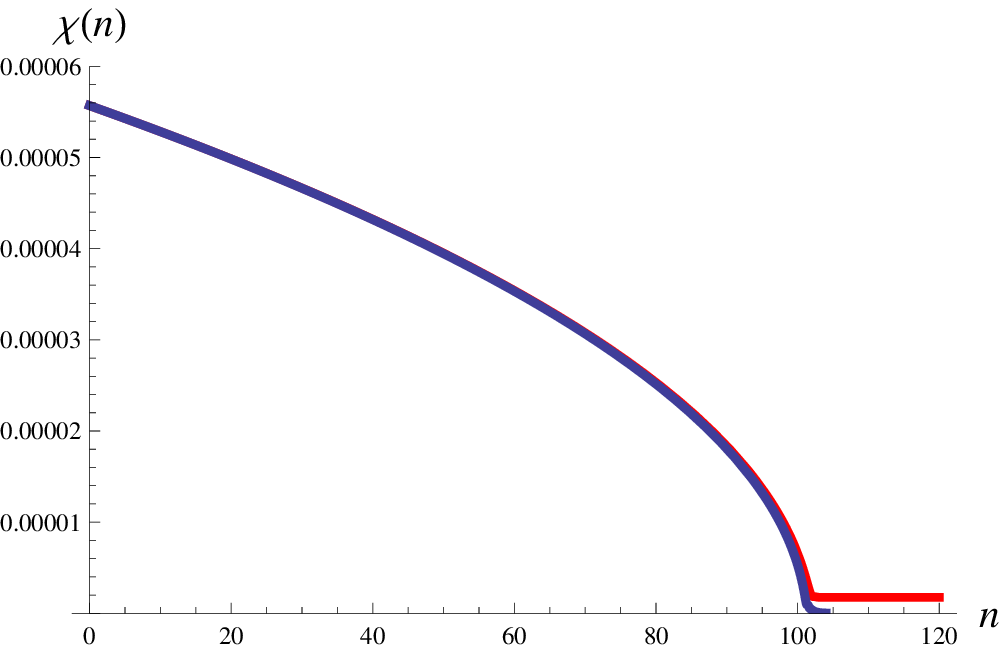}
\hspace{.5cm}
\includegraphics[width=4.0cm,height=3.0cm]{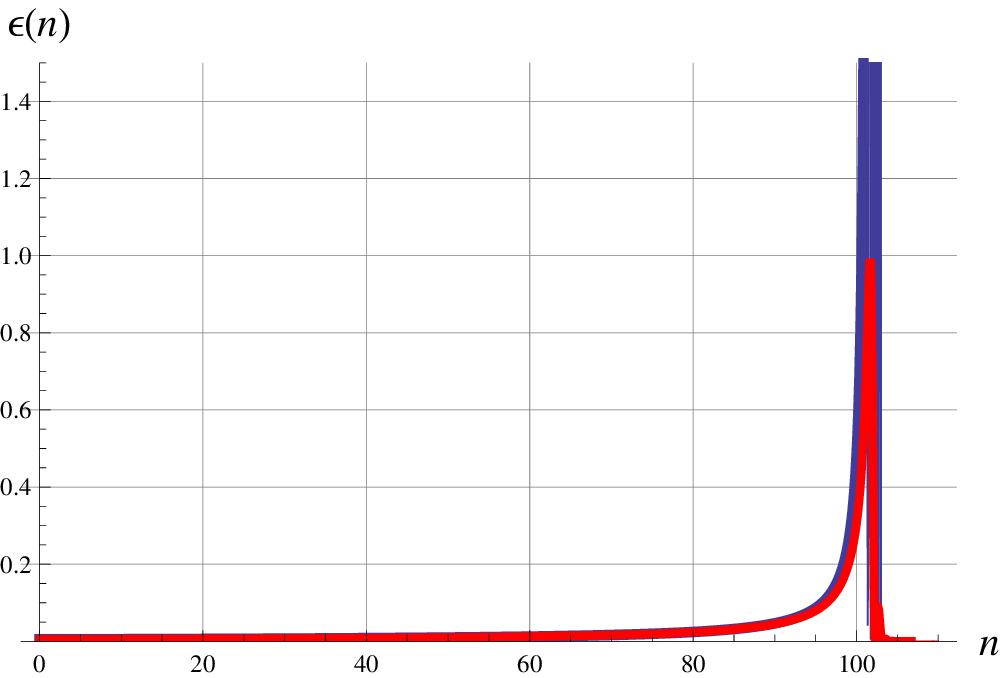}
\caption{Results from the classical model $U = \frac12 k^2 \phi^2$ (in blue) 
versus the quantum corrected model (\ref{fermiU}) (in red) assuming the inflaton
is Yukawa-coupled to a fermion. We show the dimensionless scalar $\phi(n)$ (left 
hand graph), the dimensionless Hubble parameter $\chi(n)$ (middle graph), and the 
first slow roll parameter $\epsilon(n)$ (right hand graph). The value of the 
Yukawa coupling was chosen to be $\lambda = 1.15 \times 10^{-4}$.}
\label{fermifigs1}
\end{figure}

Figure~\ref{fermifigs1} compares the quantum-corrected model (in red) with the
classical results (in blue) for the even smaller Yukawa coupling of $\lambda =
1.15 \times 10^{-4}$. Although the two models seem to track for about 100 e-foldings,
inflation ends in the classical model whereas the quantum corrected model again 
approaches de Sitter. The numerical analysis shows that $\chi$ is visibly nonzero
in this de Sitter phase whereas $\phi$ is very small.  

To see that the late de Sitter phase is generic, note that when $\phi(n)
\ll \phi(0)$ the ratio $\lambda \phi(n)/\chi(0) \ll 1$, so we can neglect the
subtraction term in (\ref{fermiU}). Now write the modified Friedmann equation 
(\ref{F1}) and the scalar evolution equation (\ref{F3}) under the assumption
that $\phi(n)$ and $\chi(n)$ are both constant,
\begin{eqnarray}
3 \chi^2 & = & \frac12 k^2 \phi^2 + \frac{\chi^4}{8 \pi^2} \Bigl[3 f(z) - z f'(z)
\Bigr] \; , \label{F1new} \\
0 & = & \frac12 k^2 \phi^2 + \frac{\chi^4}{8 \pi^2} \Bigl[ -\frac12 z f'(z)\Bigr] 
\; , \label{F3new}
\end{eqnarray} 
where $z = \lambda \phi(n)/\chi(n)$. There is no simple way to solve there equations
analytically, but it is easy to generate an efficient numerical solution. First,
subtract (\ref{F3new}) from (\ref{F1new}) to infer a relation between $\chi^2(n)$ 
and $z$,
\begin{equation}
\chi^2 = \frac{24 \pi^2}{3 f(z) \!-\! \frac12 z f'(z)} \; . \label{chiconstant}
\end{equation}
Now substitute (\ref{chiconstant}) in (\ref{F3new}) to derive an equation which
determines $z$ in terms of the parameters $k^2$ and $\lambda$,
\begin{equation}
\frac{k^2}{\lambda^2} = \frac{3 z^{-1} f'(z)}{3 f(z) \!-\! \frac12 z f'(z)} \; .
\label{zconstant}
\end{equation}
The right hand side of (\ref{zconstant}) is a complicated function of $z$ but one
can check numerically that it is monotonically decreasing. Further, the known 
asymptotic forms for $f(z)$ \cite{Miao:2015oba},
\begin{eqnarray}
{\rm Large} \; z & : & f(z) \longrightarrow z^4 \ln(z) + O(z^4) \; , 
\label{zlarge} \\
{\rm Small} \; z & : & f(z) \longrightarrow \alpha z^6 - \beta z^8 + O(z^{10}) \; ,
\label{zsmall}
\end{eqnarray}
imply that the right hand side of (\ref{zconstant}) diverges like $18 \alpha/\beta z^4$ 
for small $z$, and goes to zero like $12/z^2$ for large $z$. This means there is a
unique solution for $z$ in terms of $k^2/\lambda^2$. Hence the desired procedure is:
\begin{enumerate}
\item{Given the parameters $k$ and $\lambda$, use expression (\ref{zconstant}) to 
solve for $z$;}
\item{Substitute $z$ into (\ref{chiconstant}) to compute $\chi^2$; and}
\item{Compute $\phi^2 = z^2 \chi^2/\lambda^2$.}
\end{enumerate}

Because the late de Sitter phase emerges from numerical analysis it is no doubt 
stable. Demonstrating this analytically amounts to studying how $\frac{\partial U}{
\partial \phi}$ varies when $\phi$ is changed. Note first that altering $\phi$
induces corresponding changes in $\chi$ through relation (\ref{F1new}),
\begin{equation}
{\rm Eqn.\ (\ref{F1new})} \Longrightarrow \frac{\phi}{\chi} \frac{d \chi}{d \phi} =
\frac{-3 z f'(z) \!+\! z^2 f''(z)}{6 f(z) \!-\! 5 z f'(z) \!+\! z^2 f''(z)} \; .
\label{dchidphi}
\end{equation}
(Relation (\ref{dchidphi}) has been simplified using relation (\ref{F3new}).) A
straightforward calculation then reveals that the total derivative of 
$\frac{\partial U}{\partial \phi}$ is,
\begin{equation}
\phi^2 \frac{d}{d \phi} \Bigl( \frac{\partial U}{\partial \phi}\Bigr) = 
\frac{\chi^4}{8 \pi^2} 
\frac{[6 f (z f' \!-\! z^2 f'') \!+\! 4 (z f')^2]}{6 f \!-\! 5 z f' \!+\! z^2 f''}
\; . \label{2ndD}
\end{equation}
One can see that this is positive in the small $z$ regime (\ref{zsmall}), but not
in the regime of large $z$ (\ref{zlarge}). Because the graphs in Figures 
\ref{fermifigs2} and \ref{fermifigs1} suggest the small $z$ regime we conclude 
that the late de Sitter phase is stable.

It is not simple to derive a formula for the effective cosmological constant of
the late de Sitter phase because it depends so strongly on the dimensionless 
function $f(z)$ through relation (\ref{chiconstant}). If one assumes the small $z$ 
form (\ref{zsmall}) then the effective cosmological constant is,
\begin{equation}
\Lambda = 3 H^2 = \frac{3 \chi^2}{8\pi G} \longrightarrow 
\frac{2 \pi^2 \beta k^2}{9 \alpha^2 \lambda^4} \!\times\! m^2 \label{Lambda}
\end{equation}
Some of the numbers in relation (\ref{Lambda}) are fixed: $\alpha \simeq 0.11$,
$\beta \simeq 0.014$ and $k^2 \simeq 4.6 \times 10^{-11}$. Using the value 
$\lambda = 1.15 \times 10^{-4}$ of Figure \ref{fermifigs1} gives $\Lambda \simeq 
(7 \times 10^{5}) \times m^2$. However, our formula (\ref{Lambda}) predicts that 
decreasing $\lambda$ should {\it increase} $\Lambda$, whereas exactly the opposite 
trend is apparent in the transition from Figure \ref{fermifigs2}, with $\lambda = 
5 \times 10^{-4}$, to Figure \ref{fermifigs1}, with $\lambda = 1.15 \times 10^{-4}$. 
We attribute the apparent contradiction to the fact that ratio $k^2/\lambda^2$ is 
in neither case large enough (it is about $2 \times 10^{-4}$ for Figure 
\ref{fermifigs2} and about $3 \times 10^{-3}$ for Figure \ref{fermifigs1}) to 
justify the small $z$ approximation (\ref{zsmall}) for $f(z)$.

Finally, we consider whether the small positive cosmological constant of the
late de Sitter phase can be absorbed by adding a negative constant $-K$ to the
potential $U(\phi,\chi)$, which changes (\ref{F1new}) to,
\begin{equation}
3 \chi^2 = -K + \frac12 k^2 \phi^2 + \frac{\chi^4}{8\pi^2} \Bigl[3 f(z) \!-\!
z f'(z)\Bigr] \; . \label{F1newnew}
\end{equation}
The scalar equation (\ref{F3new}) is unchanged so relation (\ref{chiconstant})
becomes,
\begin{equation}
\chi^2 = \frac{24 \pi^2}{3 f \!-\! \frac12 z f'} \Biggl\{ \frac12 + \frac12
\sqrt{1 + \frac{K}{18\pi^2} \Bigl[3 f \!-\! \frac12 z f'\Bigr]} \Biggr\} . 
\label{chinew}
\end{equation}
And the relation which fixes $z$ changes from (\ref{zconstant}) to,
\begin{equation}
\frac{k^2}{\lambda^2} = \frac{3 z^{-1} f'(z)}{3 f(z) \!-\! \frac12 z f'(z)} 
\Biggl\{ \frac12 + \frac12 \sqrt{1 + \frac{K}{18\pi^2} \Bigl[3 f \!-\! \frac12 
z f'\Bigr]} \Biggr\} . \label{znew}
\end{equation}
Although the function on the right hand side of (\ref{znew}) still diverges as
$z \rightarrow 0$, it no longer vanishes for $z \rightarrow \infty$. Hence one
can certainly solve for $z$ when $\lambda$ is very small, but making $\lambda$
larger eventually precludes a solution. When there is a solution, its value 
will generally be larger than for $K = 0$, and this generally leads to a smaller
value of $\chi$. However, note that any value of $K > 0$ for which there is a
solution to equation (\ref{znew}) will correspond to a nonzero value of $\chi$.
So we conclude that it is only possible to avoid the late de Sitter phase by 
making $K$ large enough that (\ref{znew}) has no solution.

\subsection{Charged Inflaton Coupled to Gauge Bosons}

\begin{figure}[H]
\includegraphics[width=4.0cm,height=3.0cm]{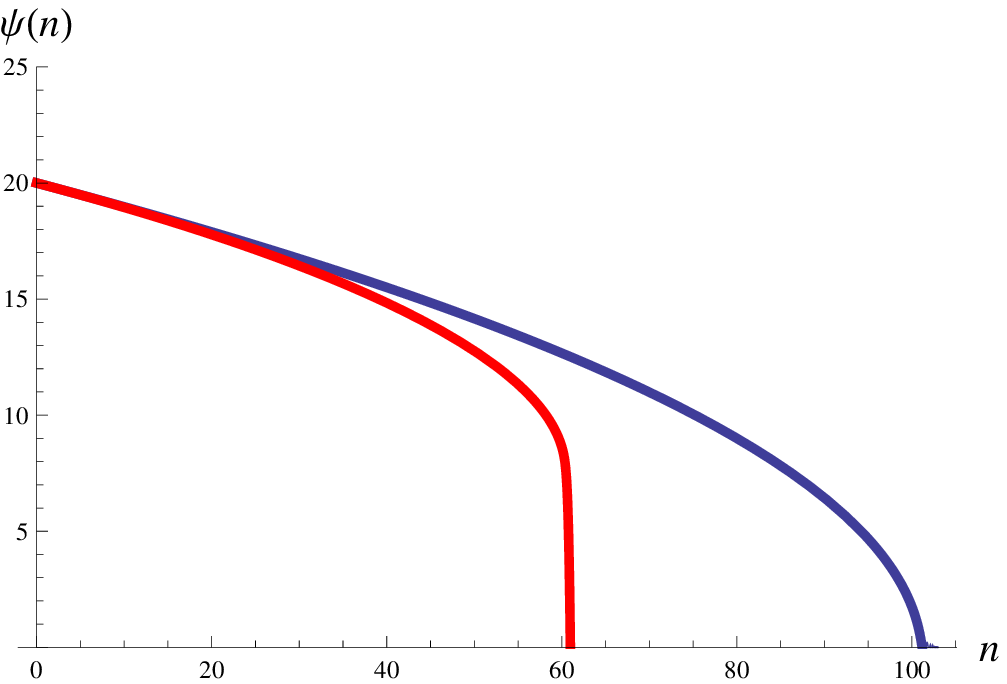}
\hspace{.5cm}
\includegraphics[width=4.0cm,height=3.0cm]{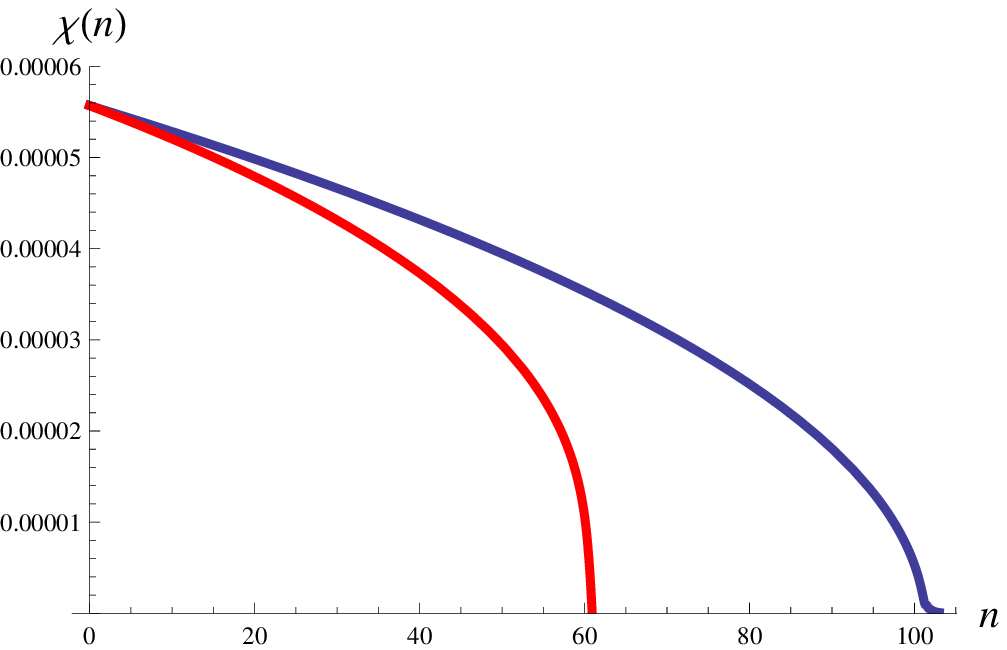}
\hspace{.5cm}
\includegraphics[width=4.0cm,height=3.0cm]{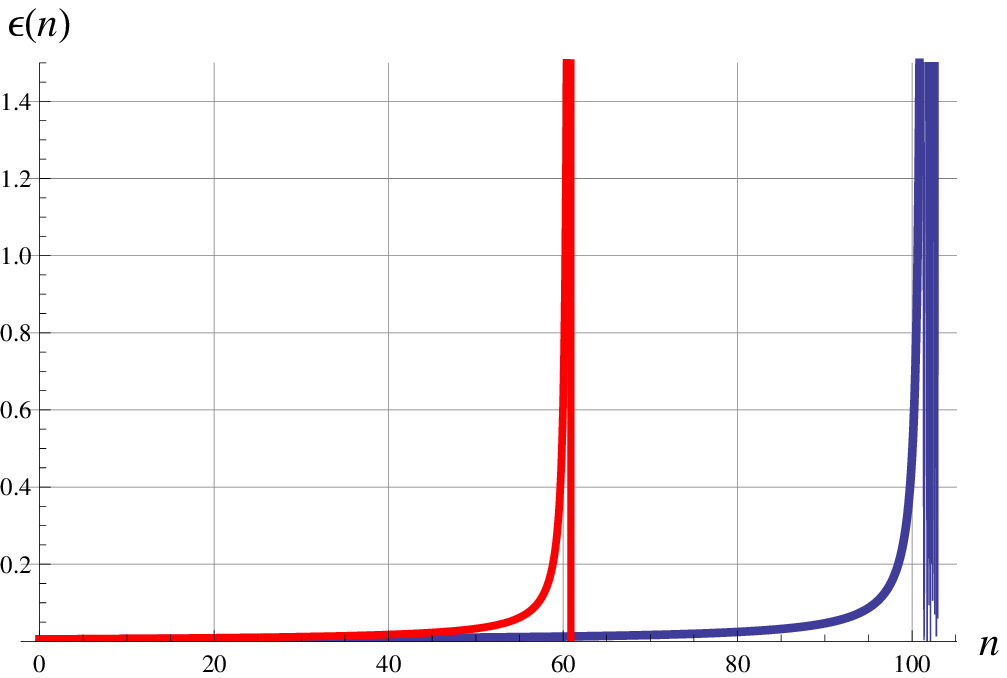}
\caption{Results from the classical model $U = \frac12 k^2 \phi^2$ (in blue) 
versus the quantum corrected model (\ref{bosonU}) (in red) assuming a charged
inflaton (with charge $q^2 = 5.5 \times 10^{-6} e^2$) is minimally coupled to 
vector bosons. The left hand graph shows the scalar $\phi(n)$, the middle graph 
gives the dimensionless Hubble parameter $\chi(n)$, and the right hand graph 
depicts the first slow roll parameter $\epsilon(n)$.}
\label{bosonfigs}
\end{figure}

The quantum-corrected dimensionless potential for a charged inflaton (with
charge $q$) is,
\begin{equation}
U(\phi,\chi) = k^2 \phi^* \phi + \frac{3 \chi^4}{8 \pi^2} f\Bigl( 
\frac{q^2 \phi^* \phi}{\chi^2}\Bigr) - \frac{3 \chi^4(0)}{8\pi^2} f\Bigl( 
\frac{q^2 \phi^* \phi}{\chi^2(0)}\Bigr) \; . \label{bosonU}
\end{equation}
The function $f(z)$ appropriate for a gauge boson is \cite{Allen:1983dg,
Prokopec:2007ak},
\begin{eqnarray}
\lefteqn{f(z) = -\Bigl(1 \!-\! 2 \gamma\Bigr) z - \Bigl( \frac32 \!-\! 
\gamma\Bigr) z^2 } \nonumber \\
& & \hspace{2cm} + \int_{0}^{z} \!\!\! dx \, (1 \!+\! x) \Biggl[ 
\psi\Bigl(\frac32 \!+\! \frac12 \sqrt{1 \!-\! 8 x} \, \Bigr) + 
\psi\Bigl(\frac32 \!-\! \frac12 \sqrt{1 \!-\! 8 x} \, \Bigr)\Biggr] \; .
\qquad \label{bosonf} 
\end{eqnarray}
Of course a bosonic quantum correction adds to the vacuum energy, which makes
the result opposite to that for fermions. For order one values of the inflaton
charge $q$ the two quantum corrections totally dominate the classical term
and inflation ends almost instantly. Making inflation last for 60 e-foldings 
requires the minuscule value of $q^2 = 5.5 \times 10^{-6} e^2$, the effects of
which are shown in Figure~\ref{bosonfigs}. Even with this charge there are 
noticeable deviations from the classical model, in particular, a much more 
sudden end to inflation.

\section{Discussion}

Scalar-driven inflation suffers from many fine-tuning problems. These are
exacerbated by the need to couple the inflaton to normal matter in order 
to make reheating efficient. Quantum fluctuations of normal matter 
induce cosmological Coleman-Weinberg potentials which are not 
Planck-suppressed and, for de Sitter, depend in complicated ways on the 
dimensionless ratio of the square of the coupling constant times the 
inflaton over the Hubble parameter. Although exact results do not exist
for more general backgrounds, it is possible to show that the factors of
``$H^2$'' are not generally constant, nor even local functionals of the
metric. The absence of locality restricts the extent to which these
corrections can be subtracted off. The purpose of this paper was to study
the consequences to inflation under two assumptions:
\begin{enumerate}
\item{The de Sitter Hubble constant is replaced by the evolving Hubble
parameter $H(t)$ in the cosmological Coleman-Weinberg potentials; and}
\item{The potentials are completely subtracted at the beginning of 
inflation with the de Sitter Hubble constant replaced by the initial
value of the Hubble parameter.}
\end{enumerate}

In section 3 we derived the appropriate generalizations to the Friedmann
equations, and we cast the formalism in terms of dimensionless variables
evolved with respect to the number of e-foldings from inflation. In 
section 4 we numerically evolved the $m^2 \varphi^2$ model, assuming
first that the inflaton is Yukawa-coupled to a fermion and then that a
charged inflaton is minimally coupled to a gauge boson. The results were
catastrophic. For the case of fermions inflation never really 
ends, no matter how small the Yukawa coupling. For bosons the 
quantum-corrected effective potential causes inflation to end almost \
instantly unless the charge is chosen so small as to make reheating
problematic.

These results are completely unacceptable for scalar-driven inflation. 
However, it is not known how much they depend upon the particular subtraction
scheme we studied. It is worth investigating subtractions based on replacing 
the factors of ``$H^2$'' by $\frac1{12} R$. That replacement would be perfect 
for the de Sitter approximation to the Coleman-Weinberg potential, but there 
is still a difference between any local subtraction and the nonlocal 
Coleman-Weinberg potential it attempts to cancel. To study this difference 
we would need a more refined analysis of the nonlocal Coleman-Weinberg 
potential. In particular, what is a generally applicable approximation for 
the de Sitter factors of ``$H^2$''? Attempting to answer this question 
seems worthwhile in view of the crippling potential problem to the viability 
of scalar-driven inflation that the current study has exposed.

Another potential solution is to couple derivatives of the inflaton to 
ordinary matter, for example $-\frac1{M^3} \partial_{\mu} \varphi \partial_{\nu} 
\varphi g^{\mu\nu} \overline{\psi} \psi \sqrt{-g}$, where $M$ is some mass 
scale. For small enough $M$ such a coupling would still be effective at 
communicating inflaton kinetic energy to the matter sector, and it has the 
virtue of preserving the (approximate) shift symmetry which is strongly 
suggested by the data. Of course the quantum corrections from such a coupling 
make no change at all in the inflaton effective potential, however, they {\it do} 
change the inflaton kinetic energy in ways that may be problematic. On de Sitter 
background the induced effective kinetic energy is closely related to the induced 
effective potential for nonderivative couplings,
\begin{eqnarray}
{\rm Nonderivative} & \Longrightarrow & -\frac{H^4}{8 \pi^2} f\Bigl(
\frac{\lambda \varphi}{H}\Bigr) \; , \label{CWpot}\\
{\rm Derivative} & \Longrightarrow & -\frac{H^4}{8 \pi^2} f\Bigl(
\frac{\partial_{\mu} \varphi \partial_{\nu} \varphi g^{\mu\nu}}{M^3 H}\Bigr) 
\; . \label{Kessence}
\end{eqnarray}
What emerges from (\ref{Kessence}) is a quantum-induced k-essence model.
Instead of order one changes in the inflaton potential we must now confront 
order one changes in the kinetic energy, which can of course alter the
inflationary geometry, the scalar and tensor power spectra and the reheat
temperature. K-essence models sometimes also permit super-luminal propagation.
It would be fascinating to make a quantitative study of the various
consequences.

\vskip 1cm

\centerline{\bf Acknowledgements}

This work was partially supported by Taiwan MOST grants 
103-2112-M-006-001-MY3 and 106-2112-M-006-008-; by NSF grants PHY-1506513 and 
PHY-1806218; and by the Institute for Fundamental Theory at the University of 
Florida.

\end{document}